\documentstyle[eqsecnum,aps,preprint]{revtex}
\tightenlines
\def\mathcalH{{\cal H}}
\def\bm{\boldmath}
\begin{document}
\draft
\title{Vortex dissipation and level dynamics for the layered superconductors
with impurities}
\author{Ayumi Fujita}
\address{Center for Promotion of Computational Science and Engineering (CCSE),
Japan Atomic Energy Research Institute, 
2-2-54 Nakameguro, Meguro-ku, Tokyo 153-0061, Japan
}
\date{\today}
\maketitle
\begin{abstract}
We study parametric level statistics of the discretized excitation 
spectra inside a moving vortex core in layered superconductors with impurities.
The universal conductivity is evaluated numerically for the various values 
of rescaled vortex velocities $\kappa$ from the clean case to the dirty
limit case.  The random matrix theoretical prediction is verified 
numerically in the  large $\kappa$ regime.
On the contrary in the low velocity regime, we observe 
$\sigma_{xx} \propto \kappa^{2/3}$
which is consistent with the theoretical result for the super-clean case,
where the energy dissipation is due to the Landau-Zener transition which
takes place at the points called ``avoided crossing''.
\end{abstract}

\section{Introduction}
The discrete energy spectra inside the vortex core in 
two-dimensional superconductors play an important role for  
transport properties in the mixed state especially in low temperatures.  
The energy dissipation due to the vortex motion results from the transition
between these levels.
The $I$-$V$ characteristic in superconductors has been studied by many
authors.  In the superclean case $\omega_0 \tau \gg 1$, where $\omega_0$
is the level separation and $\tau^{-1}$ is the inverse scattering time,
the dissipative conductivity has been obtained quasiclassically as
$\sigma_{xx} \simeq en_e /(B\omega_0 \tau)$.~\cite{kopnin}
For the 2D superconductor in the superclean limit where there is only one 
impurity inside the vortex core,  the energy dissipation due to the 
Landau-Zener transition between levels has been evaluated.
The energy transfer takes place when the impurity passes through 
the ``dangerous region'', where many Landau-Zener transitions 
contributing to the dissipation.~\cite{LO} 
There is also a perturbational approach for the moving
impurity which is valid in the high velocity limit $v \gg v_F (\Delta/E_F
)^2$ where $v$ is the vortex velocity and $\Delta$ is the mean
level spacing.~\cite{Guinea}  
The transition between time-dependent levels $E_i (X)$,
where $X$ is the external perturbation (the position of impurities)
are treated in ref.\onlinecite{Feigel} in the framework of random
matrix theory, and almost the same result for
$\sigma_{xx}$ as the quasi-classical one is obtained.

The parametric level statistics for the energy
spectra under various external perturbations provide a new 
perspective of transport properties in disordered quantum
systems such as metallic grains or quantum dots. The Thouless formula
which relates the conductance to an equilibrium quantity such as level 
curvatures gave the first clue to investigations of universal behaviors 
of level statistics.~\cite{thoule1,thoule2}  The random matrix
theory (RMT) and its classification of distributions of 
level spacing (Wigner-Dyson statistics),
which plays a role of a finger-print also for quantum chaotic systems, 
supplies a powerful tool for the study of disordered systems in the
framework of level dynamics.~\cite{mehta,haake}

There have been both theoretical and numerical studies within the
framework of the RMT for the level dynamics of the disordered systems
such as an ring subject to the Aharonov-Bohm flux 
and non-interacting electrons in background potentials.
~\cite{akkermans,Alts1,Alts2,Fyodorov}
For the quantum chaotic systems, there are experimental 
studies of the microwave billiards where the Gaussian distribution of 
level velocities has been observed.~\cite{Barth}
The motivation of the current study is to investigate the universal behavior 
of the level velocities for low energy excitation spectra inside a moving
vortex core dragged through a layered superconductor over a wide region
from moderately clean to dirty phase.
In the presence of disorder, the Hamiltonian for the scattering quasi-particles
between the excitation levels inside the vortex core becomes the Lie algebra
$Sp(2N)$, and it belongs to the class C of the classification of the 
Atland-Zirnbauer.\cite{AZ}
The level statistics over the crossover region 
from the moderately clean to the dirty phase has been 
investigated in the framework of the RMT~\cite{Hikami}, 
and also verified numerically
for various impurity concentration inside the core. For the energy
levels far from the Fermi energy $E_i \gg E_F$ in the dirty limit case,  
the level statistics have been shown to be identical with that of the 
Gaussian unitary ensemble (GUE).~\cite{af}

In section II, we investigate the universal distribution of the level
velocities for the quasiparticle spectra within the moving vortex core
for different regimes which are characterized by the parameter
$\omega_0 \tau$ (or the number of impurities within the core $N_i$),
with constant vortex velocity $v$.
We observe that in the case of $\omega_0 \tau \ll 1$ 
the Gaussian distribution of level velocities is observed.
In the clean case, non-Gaussian distribution with an enhanced
peak at $v=0$ which is the characteristic of the systems with
localized wave-functions~\cite{Fyodorov} is obtained.

In the diffusive time-evolution of energy spectrum within the core, 
the vortex velocity $v$ plays an essential role. 
It should be large enough that pinning is negligible, since $v$ is
taken as a constant.
However it is assumed to be moderately small
$v \ll v_K$ in order to satisfy the adiabaticity of transitions between levels.
The parameter $v_K$ is determined by the elastic mean free path and it
manifests the mesoscopic nature of the level dynamics.~\cite{Feigel}

In section III, the energy dissipation for the low velocity regime is
investigated and nonlinear $I$-$V$ characteristics are studied.
We follow the same procedure as in ref.\onlinecite{wil} in which  
the energy dissipation has been evaluated through iterative calculations 
of time-dependent spectra from the Schr\"odinger equation for a 
perturbed Hamiltonian.  Our result coincides with that obtained for the
superclean case in ref.\onlinecite{LO} in the small $v$ regime.  
In the dirty phase ($\omega_0 \tau \ll 1$), our numerical result is 
identical to that of the random matrix theory for the Gaussian unitary 
ensemble with moderately large $v$.

\section{Universality in level velocity distribution}

The excitation spectrum inside the vortex core is obtained from the
Bogoliubov-de Gennes equation,~\cite{caroli}
\begin{equation}
\hat{\mathcalH}
{u\choose v} =
E {u\choose v},
\end{equation}
where
\begin{equation}
\hat{\mathcalH}=\left(
\begin{array}{cc}
{\bm p}^2 /2m-E_F +V({\bm r}) & \Delta({\bm r}) \cr
\Delta^\ast ({\bm r})       & -{\bm p}^2 /2m+E_F -V({\bm r})
\end{array}\right).
\end{equation}
We consider the case for the $s$-wave gap parameter
\begin{equation}
\Delta({\bm r})=\Delta(r)e^{i\phi},
\end{equation}
where $\phi$ is the polar angle.
With the absence of impurity ($V({\bm r})=0$), the excitation energy spectrum
\begin{equation}
E_n^0 = -(n-\frac{1}{2})\hbar\omega_0 ,\qquad n=0,\pm1,\pm2,\cdots
\end{equation}
is obtained.  We set $\hbar=1$ hereafter and 

\begin{eqnarray}
\omega_0 & =&\int_0^\infty [\Delta(r)dr/k_F r]e^{-2K(r)}\bigg/
\int_0^\infty dr e^{-2K(r)}, \\
K(r) & =&\int_0^r dr' \Delta(r')/v_F .
\end{eqnarray}

If we take $\Delta(r)=\Delta \quad(=\Delta_\infty)$ 
then the level spacing becomes
$\omega_0 \sim \Delta^2 /E_F$.  The $n$-th eigenfunction is
\begin{equation}
\hat\phi_n =
{u_n\choose v_n}
=\tilde c e^{-K(r)}\left(\begin{array}{c}e^{in\phi}J_n (k_F r)\\
-e^{i(n-1)\phi
}J_{n-1}(k_F r)\end{array}\right),
\end{equation}
which we call Caroli-de Gennes-Matricon (CDM) basis where $J_n (x)$ denotes a
Bessel function and $\tilde c$ is a normalization constant.

When the impurity scattering is present, the wave function
becomes a linear combination of the CDM basis as 
\begin{equation}
\hat\psi_n =\sum_m \hat f_{mn} \hat\phi_m 
\end{equation}
where $n$-th column of the matrix $\hat f$ is given by the first order 
perturbation as
\begin{equation}
\label{eq:first}
(\hat\varepsilon -\hat E +\sum_{i}\hat A^i )\hat f_n =0,
\end{equation}
\begin{equation}
\hat\varepsilon_{nm}=\delta_{nm}E_n^0, 
\end{equation}
\begin{equation}
\hat A^i_{nm}=V_i e^{-2K(r_i )+i\phi_i (m-n)}[J_n (k_F r_i )J_m (k_F r_i )
-J_{n-1}(k_F r_i )J_{m-1}(k_F r_i )]/\lambda_F \xi,
\end{equation}
with the assumption that impurity potential $V({\bm r})$ to be a sum of 
short-range scattering sites ${\bm r}_i =(r_i , \phi_i )$
\begin{equation}
V({\bm r})=\sum_i V_i \delta({\bm r}-{\bm r}_i ).
\end{equation}
The summation for $\hat A^i$ in eq.(\ref{eq:first}) is taken for all 
impurity sites and $\xi$ is the coherence length.

When the vortex is dragged within the sample due to the Lorenz force,
the Hamiltonian
\begin{equation}
\label{eq:hamil}
\hat {\mathcalH}(X(t)) = \hat\varepsilon + \sum_i \hat A^i 
\end{equation}
is perturbed by changing realizations of the impurity positions
within the core.
The perturbation parameter $X(t)$ is the vortex coordinate and the
energy levels go up and down at different times (fig.1). Then
the Hamiltonian becomes time dependent with
the vortex velocity $v$ fixed as a constant.

The universal Gaussian distribution of the level velocities was 
studied for disordered metallic rings subject to the Aharonov-Bohm flux.
~\cite{Alts2}
The level dynamics become universal dependent only on the Wigner-Dyson ensemble
after a rescaling of the energy $E$ and the controlling parameter $X$ as
\begin{eqnarray}
\varepsilon(x) &=& E(X)/\Delta, \\
x &=& \sqrt{C(0)} X,
\end{eqnarray}
where $\Delta$ is the mean level spacing and $C(0)$ is the
normalized dispersion of level velocities,
\begin{equation}
C(0)={\frac{1}{\Delta^2}}\left< \left(\frac{dE_i}{dX}\right)^2\right>.
\label{c0}
\end{equation}
The statistical average $<\cdots>$ is taken over samples and 
$X$, i.e. over time steps as we fix the vortex velocity $v$.  
It was shown that the equilibrium quantity $C(0)$ is directly related to
the transport (dissipative) conductivity given by Kubo formula.~\cite{akkermans}
Thus $C(0)$ is also called the universal conductance.

The density of states of quasi-particles inside the vortex core in dirty or
moderately clean phase has been investigated theoretically.
~\cite{Skvortsov,Koulakov,Hikami}
It was also shown from the explicit diagonalization of the Hamiltonian 
matrix that in the dirty limit the Wigner-Dyson statistics of level 
distributions become
identical to that of the Gaussian unitary ensemble when the levels are
far from the Fermi energy.~\cite{af}
Following the same procedure we evaluate the distribution of the rescaled 
level velocities $P(\dot\varepsilon_i )$ where $\dot\varepsilon_i$
is written as
\begin{equation}
\dot\varepsilon_i = \frac{dE_i}{dt}\cdot\frac{1}{\Delta\sqrt{C(0)}},
\end{equation}
for various numbers of impurities $N_i$ within the vortex core.

The inverse scattering time is given by
\begin{equation}
\tau^{-1} =2n_i V_i^2 m.
\end{equation}
$n_i =N_i /\xi^2$ is the concentration of the impurities within the
core of radius $\xi$, where $\xi$ is the coherence length.
In the numerical calculations we set $V_i =1$ and $\xi k_F =50$.
Thus from the relation $\xi\sim k_F /m\Delta$ we have
\begin{equation}
(\omega_0 \tau)^{-1}=N_i .
\end{equation}
The size of the Hamiltonian matrix $\hat{\mathcalH}$ is taken as 
$100\times 100$.  
The vortex motion is restricted in $y$-direction and periodicity in this
direction for the configuration of moving impurities is assumed.
The discretized shift $\Delta y$ of an impurity at discretized time step
$\Delta t$ is $\Delta y = \xi v\Delta t$.
The rescaled level velocity $\dot\varepsilon_i$ for $100$ time steps are 
calculated
and the $P(\dot \varepsilon_i )$ for levels between $E_{10}$ to $E_{40}$ 
are averaged, where we denote $E_0$ for the Fermi Level ($E_0 \equiv E_F$).  
The sample averaging has 
been taken over $320$ ($480$ for $N_i =500$ case) different initial 
configurations of the impurities.

In fig.2 we show averaged distribution of the level velocities $P(\dot
\varepsilon)$ for various numbers of impurities within the core. 
We set $v=0.002$ and $\Delta t=1$.
The inset shows the rescaled variance $C(0)$ for each energy level.
The condition for the dirty phase $1\ll (\lambda_F /\xi)^2 NN_i \ll N$ where
$N$ is the half of the size of the Hamiltonian is well satisfied in the
case $N_i =500$ $(\omega_0 \tau =0.002)$. 
In this dirty phase the Hamiltonian can be described as
the random matrix and the Gaussian distribution of the level velocities
are observed.

In the clean case $N_i =3$ $(\omega_0 \tau \simeq 0.33)$, 
the non-Gaussian behavior is found.
We obtain a sharp peak at $\dot\varepsilon =0$ which is similar to 
the result of the
microwave experiment for the case when the local level dynamics are concerned.
~\cite{Barth}
This non-Gaussian behavior suggests that the wave functions for the scattering
quasiparticles in this regime is still localized.~\cite{Fyodorov}

\section{Dissipative property of the wave function
of quasi-particles inside a vortex core}

We evaluate the evolution of the wave function $|\psi(t)\rangle
=\{|\psi_1 (t)\rangle,|\psi_2 (t)\rangle,\dots, |\psi_{2N} (t)\rangle\}$ from 
the time dependent Schr\"odinger equation 
\begin{equation}
i\frac{d}{dt} |\psi(t)\rangle = \hat {\mathcalH} |\psi(t)\rangle,
\end{equation}
where
\begin{equation}
\hat {\mathcalH} = \hat\varepsilon + \sum_i \hat A^i 
\end{equation}
or equivalently consider time evolution of the coefficient matrix $\hat f$ 
which depends on time. 
\begin{equation}
\label{eq:TS}
i\frac{d}{dt}\hat f(t) = \hat {\mathcalH}\hat f(t)
\end{equation}
The initial state $\hat f(0)$ is assumed to be the $i$-th eigenvector
of $\hat{\mathcalH}$ which corresponds to the $i$-th energy level $E_i$.
\begin{equation}
\hat f(0)=\left\{ 0, \dots ,\hat f_i ,\dots,0\right\}
\end{equation}
Then the initial wave functions are
\begin{eqnarray}
|\psi(0)\rangle &=\left\{0,\dots,|\psi_i \rangle,\dots,0\right\}\nonumber\\
|\psi_i \rangle &=\sum_m \delta_{ni}\hat f_{mn}\hat\phi_m
\end{eqnarray}

When the vortex is moving, the impurity sites ${\bm r}_i$ within the core
are modified and the Hamiltonian $\hat{\mathcalH}$ becomes time dependent.   
We investigate the time evolution of $|\psi(t)\rangle$ which is expressed
in the basis $\hat g$ of the instantaneous Hamiltonian $\hat{\mathcalH}(t)$ :
\begin{eqnarray}
\hat{\mathcalH}(t)\hat g_n (t)&= E_n (t)\hat g_n (t)\\
|\tilde\psi_n (t)\rangle &=\sum_m \hat g_{mn} \hat\phi_m \\
\hat f_n &= \sum_m \hat a_{mn}(t)\hat g_m  \\
\hat a_{mn}(t) &= \hat g_m \cdot \hat f_n
\end{eqnarray}

As the vortex moves at a constant velocity $v$, the time-dependent spectrum 
shows diffusive time-evolution. 
This is the characteristic behavior which suggests that consecutive Landau-Zener
transitions occur incoherently.  
The energy transfer takes place at the points called avoided crossings,
where the separation between levels is much less than the mean level spacing $\Delta$.
We assume that the quantum interference term in the transition probability
between the consecutive levels can be ignored.  The full quantum coherent time-evolution,
however, leads to localization in energy space.
The condition for this assumption is that many Landau-Zener transitions contribute 
to the diffusion, so we may take enough statistical average.~\cite{wil}

Thus the occupation probability $|\hat \psi_n (t)|^2$ spreads diffusively as
time.  Following ref.\onlinecite{wil}, we define the probability 
distribution $\Pi_i (t)$ as
\begin{equation}
\Pi_i (t)=\sum_k (\hat g_k \cdot \hat f_k) (k-i)^2
\end{equation}
$\Pi_i (t)$ is averaged over samples (over different configurations of initial
impurity sites), and/or over many initial states $i$.  
The diffusive behavior of $<\Pi (t) >$ for small $t$ is expected to be 
linearly dependent on time,
\begin{equation}
\left< \Pi(t) \right> = R \, t.
\end{equation}
Solving the time dependent Schr\"odinger equation numerically, we
investigate the dependence of the diffusion constant $R$ on the rate of 
change of the external perturbation, i.e. on vortex velocity $v$.  

The diffusive behavior is also dependent on the smoothed density of states
$\rho$.  In this study we assume that $\rho$ is constant as $\rho=\Delta^{-1}$
where $\Delta$ is the mean level spacing.  
There are two different regimes which are characterized by the small $v$ or
the large $v$ respectively.  The calculation of the diffusion constant
$R$ for the levels far from the Fermi energy reveals the 
nonlinear behavior of the energy dissipation of quasi-particles inside the
vortex core in the low $v$ regime.

In fig.3 we plot the result for the clean phase $N_i =3$ and the dirty phase
$N_i =200$ for various values of 
$v=0.001,0.002, 0.005,0.01,0.02,0.05,0.1$ and $0.2$.  The initial state is 
$E_i =E_{35}$ and $E_i = E_{10}$, respectively near the hump of $C(0)$ plot
for $v=1$ case which is shown by the insets.
The horizontal axis is the rescaled vortex velocity $\kappa = \sqrt{C(0)}v$.
In the calculation of the numerical solution of eq.(\ref{eq:TS}), the 
discrete time step is taken as $\Delta t=0.002$ and the sample averaging has 
been done over $32$ different initial configurations of the impurities.  
The gradient of the dotted line which is the least square fit is $0.67$.
We find that $R\propto \kappa^\alpha$ where $\alpha\sim 2/3$ for 
small $v$ from the fit.
Since the vortex velocity is written as $v=E/B$, this result is consistent 
with the behavior of the dissipative part of the 
current density $j_x$ which is obtained in ref.\onlinecite{LO} as
\begin{equation}
j_x ={{a_0 n_{\text{imp}}}\over {\phi_0}}{{E_F^{5/3}}\over {\Delta^{2/3}}}
\left({E \over {v_F B}}\right)^{2/3},
\end{equation}
where $a_0$ is the distance from the "dangerous region" to the vortex center
and $\phi_0 =\pi/e$ is the flux quantum. 
This formula is valid for $v\ll v_F (\Delta/E_F )^2$ and in the 
super-clean limit $N_i =1$ case. Since $C(0)$ is suppressed with
the decrease of $\omega_0 \tau$ (or increase of $N_i$), the random matrix
prediction is irrelevant as well as in the large $N_i$ case 
for this low $v$ regime as expected. 
We also note that for larger values of $v$,
we obtain larger values of $\alpha\sim 0.9$ in all dirty cases of 
$N_i =50, 100$ and $200$ for which we have done the calculation.

In the meanwhile for the large $v$ regime the random matrix theory predicts 
the diffusion constant $R$ for the perturbed Hamiltonian whose matrix
elements are taken from 
the Gaussian unitary ensemble (this is the relevant for the current study) as 
\begin{equation}
R = \pi C(0) \dot X^2
\end{equation}
where $C(0)$ is given by eq.(\ref{c0}) and $\dot X = v$.
The vortex damping coefficient $\eta$ is defined as
\begin{equation}
\partial W/\partial t = \eta v^2
\end{equation}
where l.h.s is the rate of the energy dissipation inside the core and $\eta$ is
related to the dissipative conductivity as $\sigma_{xx}=\eta ec/\pi B$.
From the quasi-classical result in the $\omega_0 \tau \ll 1$ case, the damping
coefficient $\eta$ is given by~\cite{kopnin}
\begin{equation}
\eta = \pi n_{2D}\omega_0 \tau
\end{equation}
where $n_{2D}$ is the density of electrons.  Thus the random matrix theory 
predicts~\cite{Feigel}
\begin{equation}
C(0)=n_{2D} (\omega_0 \tau).
\end{equation}
In fig.4 we plot $C(0)$ v.s. $\omega_0 \tau$ for various $v$ values.  
In the large $v$ case the relation $C(0)\propto \omega_0 \tau$ is explicitely 
observed.

\section{Discussion}
In this study we have investigated numerically for the universal behavior of the
distribution of the rescaled level velocities for the excitation
spectra within the moving vortex core in layered superconductors.  
In the dirty phase where $\omega_0 \tau \ll 1$ we obtain the Gaussian 
distribution consistent with the result of the theoretical study for the 
nonlinear $\sigma$ model in the various disordered quantum systems.
Contrary to this,  the non-Gaussian behavior which is considered as the 
characteristic
property which manifests the localization of wave functions inside the core
is observed within the clean phase $\omega_0 \tau \sim 1$.

As to the energy dissipation within the core due to the Landau-Zener transition
between levels which occurs at the points called ``avoided crossings'',
the nonlinear dependence of the $\sigma_{xx}$ on the external
perturbation (the vortex velocity) is investigated.  This is explicitly related
to the nonlinear $I$-$V$ characteristics in the mixed state
in low temperature region.  

The dissipation properties are classified into
three different regimes characterized by two parameters, the number of
impurities inside the core $N_i$ (or equivalently the inverse of $\omega_0 
\tau$), and the vortex velocity $v$. Case i)Large $v$ and large $N_i$ regime; 
in this regime the prediction of the RMT is 
valid and at the same time the behavior becomes consistent with those
obtained from the approach within the Kubo-Greenwood formula. Case ii)Small 
$v$ and large $N_i$ regime; in this regime the nonlinear behavior of the 
$\sigma_{xx}$ is observed.  Since the variance of 
the scaled level velocity $C(0)$ is suppressed as the increase of $N_i$, the
random matrix theoretical result is irreverent in this regime no matter how
$N_i$ becomes large.  
However, we consider that a
crossover behavior for the dissipation properties within the core should
exist in this regime, which has been found both theoretically and 
numerically in the case of the density of states.  Thus the third regime
should be the crossover regime denoted as 
iii)Large $N_i$ and moderately large $v$ regime. We consider
the result of large $\alpha$ values which are obtained in fig. 4 in large
$v$ case implicitly manifest this crossover behavior.
The crossover properties in this regime
from the super-clean result of Larkin-Ovchinikov \cite{LO} to the dirty 
case where RMT result should become valid is the issue of future study.

\acknowledgements
The numerical calculations has been done on the HITACHI 
SR2201 with 48 processors at CCSE.

\begin{figure}
\caption{A snap shot of the perturbed spectra inside a vortex core where the external
perturbation is due to the motion of the vortex at velocity $v$. The number
of impurities inside the core is $N_i =3$. The discretized time step is $\Delta t=0.06$
and $v=0.1$. The size of the Hamiltonian
matrix is $100\times 100$ and $10$ levels above and below the Fermi level 
$E_i =0$ are shown.\label{fig1}
}
\end{figure}

\begin{figure}
\caption{The averaged distribution of the level velocities 
$P(\dot \varepsilon_i )$ for the vortex velocity $v=0.002$ 
with various numbers of impurities $N_i$ inside the core.
The solid line corresponds to the Gaussian distribution.
The inset shows the dispersion of level velocities $C(0)$. \label{fig2}
}
\end{figure}

\begin{figure}
\caption{Plot of dissipation constant $R$ versus the normalized vortex 
velocity $\kappa$ for ${N_i} =3$ and ${N_i} =200$ cases.
The result for ${N_i} = 200$ is shifted downwards in order to see the coincidence
of the slope of the fitting lines in lower $\kappa$ regime.
The time step is taken as $\Delta t=0.002$.
The gradient of the dotted line is $\alpha=0.67$ (small $\kappa$) and
the dashed line $\alpha=0.89$ (large $\kappa$).\label{fig3}
}
\end{figure}

\begin{figure}
\caption{Graph of log-log plot of $C(0)$ vs $\omega_0 \tau$ ($=1/N_i$) 
for various values of vortex velocity $v$ for $\Delta t=1$.\label{fig4}
}
\end{figure}

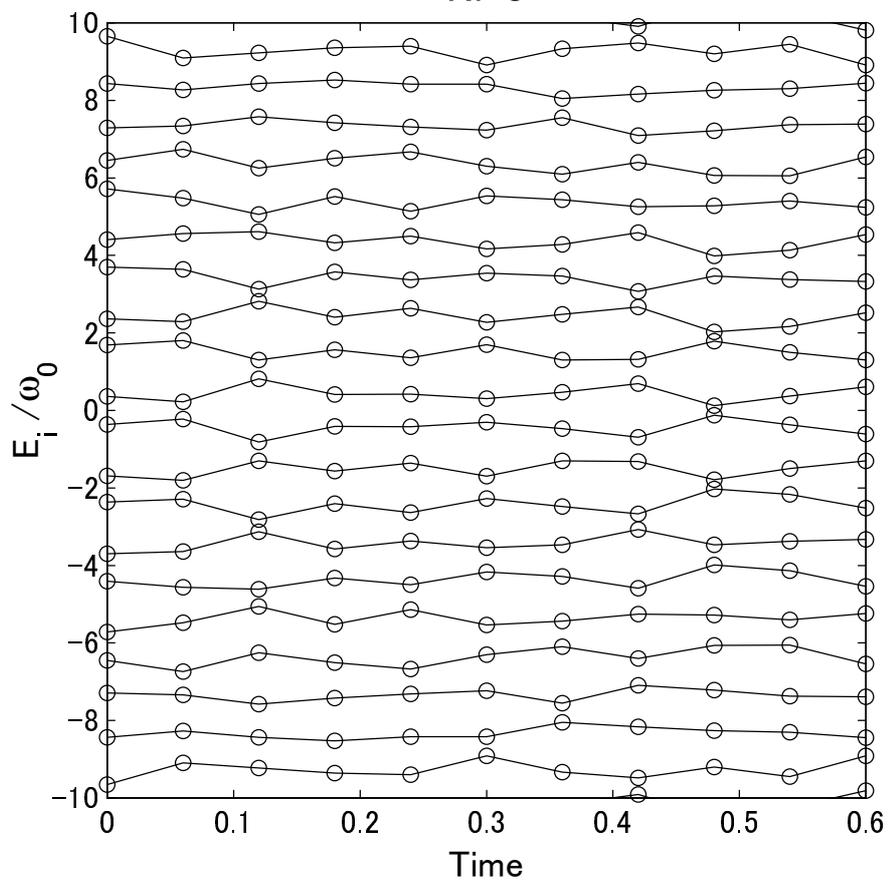
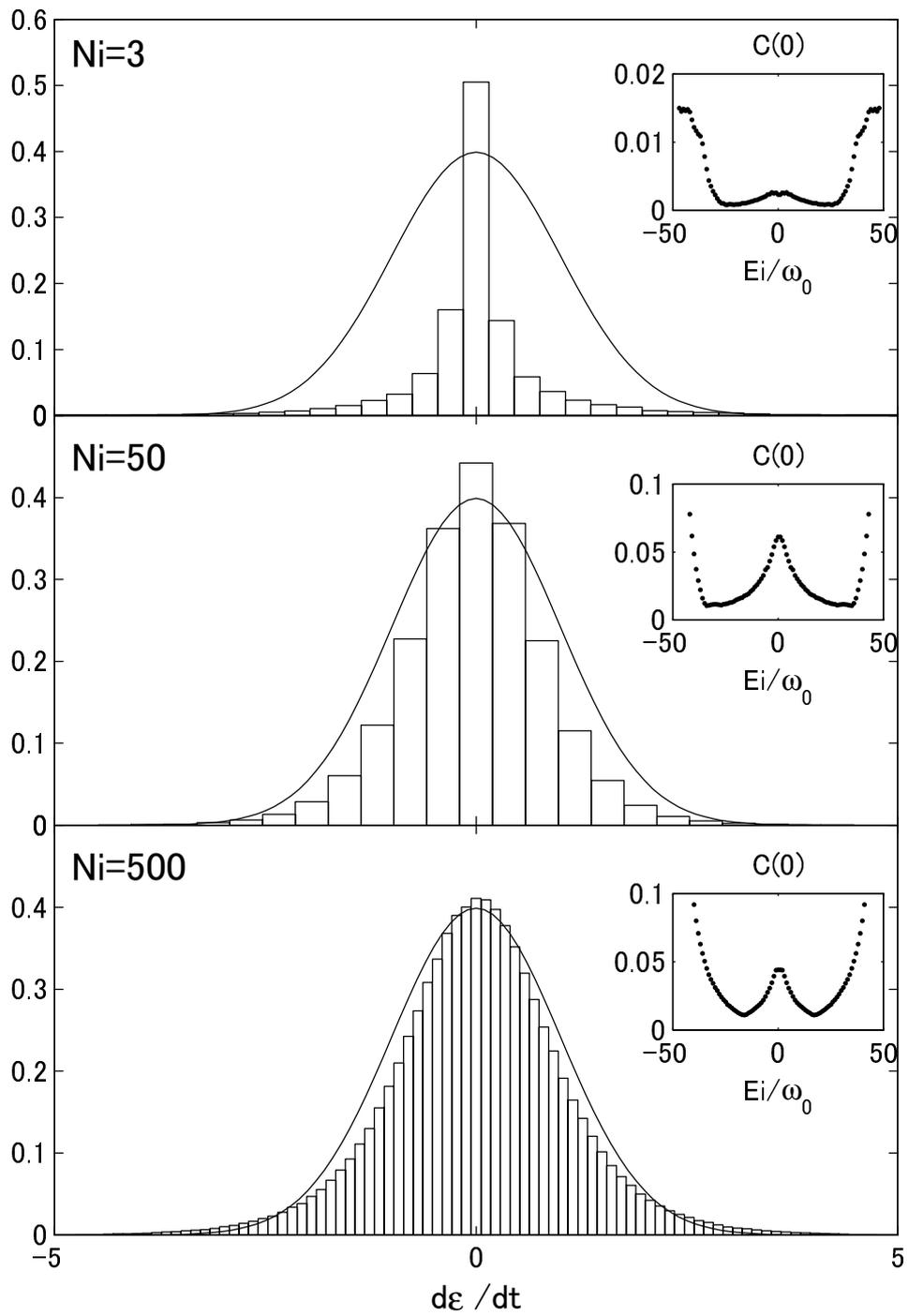
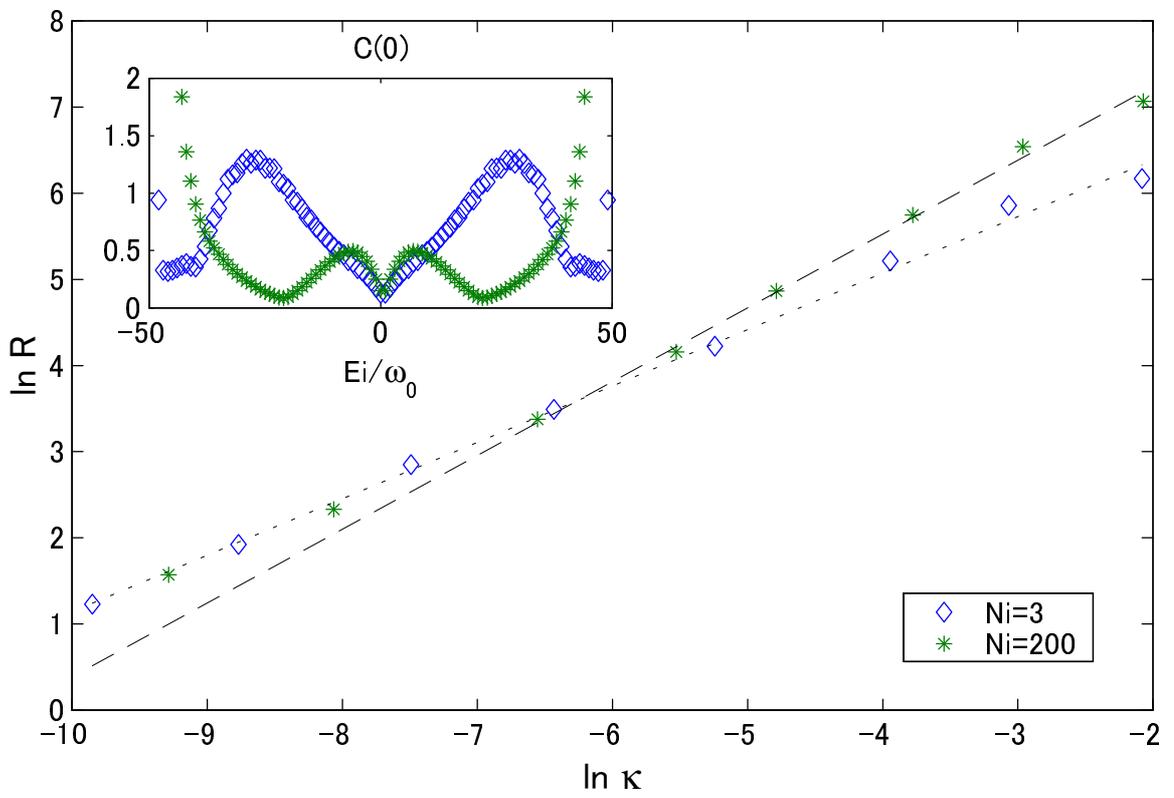
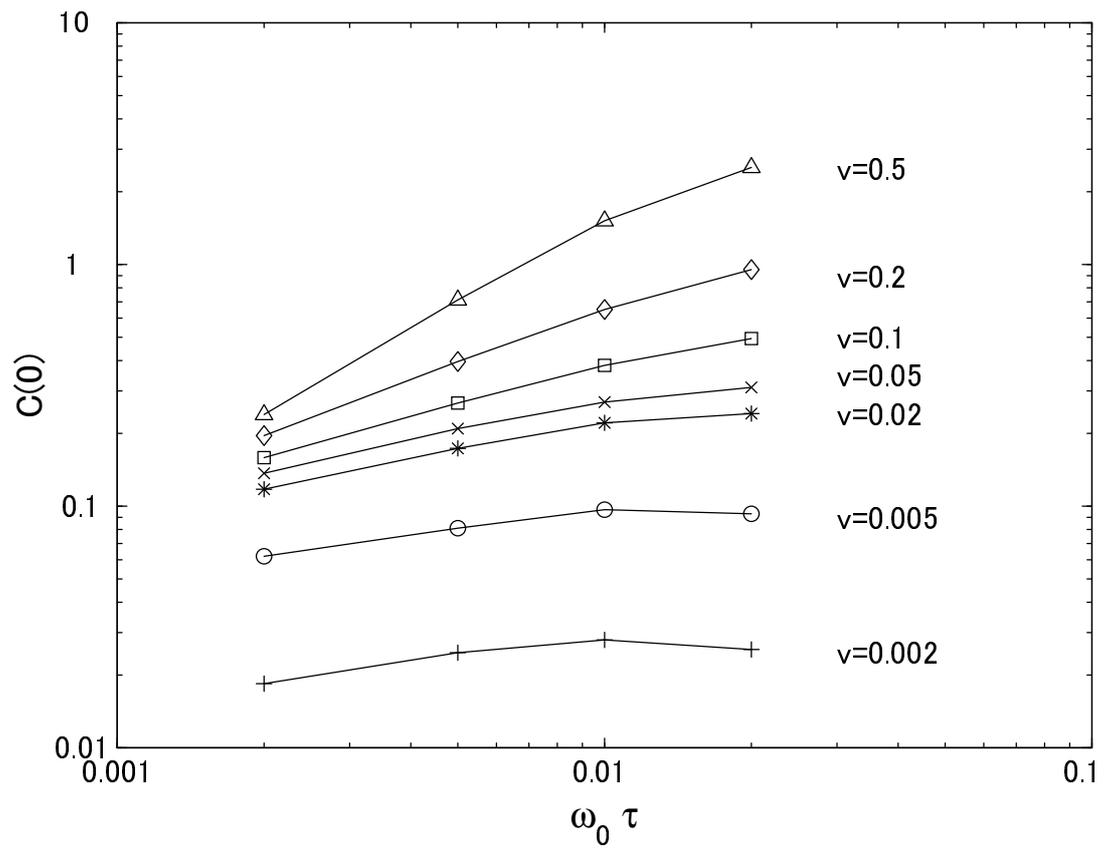
\end{document}